\documentclass[twocolumn,prl,aps]{revtex4}
\usepackage{floatflt}
\usepackage{psfrag}
\usepackage{amsmath}   
\usepackage{amssymb}
\usepackage{amsfonts}
\usepackage{mathrsfs}
\usepackage{graphicx}
\usepackage{xcolor}
\usepackage{physics}

\begin{document}
\title
{Plasmons in a Square of Two-Dimensional Electrons}
\author{A.~M.~Zarezin$^{a,b}$, D.~Mylnikov$^{b}$, A.~S.~Petrov$^{b}$, D.~Svintsov$^{b}$, 
P.~A.~Gusikhin$^{a}$, I.~V.~Kukushkin$^{a}$, V.~M.~Muravev$^{a}$ \footnote{Corresponding E-mail: muravev@issp.ac.ru}}
\affiliation{$^a$ Institute of Solid State Physics, RAS, Chernogolovka, 142432 Russia \\
$^b$ Moscow Institute of Physics and Technology, Dolgoprudny 141700, Russia}
\date{\today}
\date{\today}

\begin{abstract}
Microwave absorption spectra of a single square of two-dimensional electrons (2DES) have been investigated using an optical detection technique. Fundamental dipole and harmonic quadrupole plasmon modes have been identified and compared to those in the disk geometry. In the square-shaped 2DES, a strong interaction is discovered between the neighboring plasmon modes, whereas no such hybridization is observed in the disk-shaped geometry. We establish a rigid theoretical platform to analytically describe the magneto-optical response of confined two-dimensional systems. The developed theory provides a proper description of the obtained experimental results.
\end{abstract}

\maketitle

\section{Introduction}

Plasma excitations in two-dimensional electron systems (2DES) have been the focus of active research in recent years~\cite{Andress:2012, Koppens:2012, Basov:2012, Aizin:2013, Lusakowski:2016, Zudov:2016}. The spectrum of two-dimensional (2D) plasmons has been well established both theoretically~\cite{Stern:1967} and experimentally~\cite{Grimes:1976, Allen:1977, Theis:1977} as follows: 
\begin{equation}
\omega_p (q)=\sqrt{\frac{n_s e^2 q}{2m^{\ast} \varepsilon_0  \varepsilon (q)}}  \qquad (q \gg \omega/c).
\label{2D}
\end{equation}
Here, $q$ is the wave vector of the plasmon, $n_s$ and $m^{\ast}$ are the density and the effective mass of the 2DES electrons, while $\varepsilon_0$ and $\varepsilon (q)$ denote the vacuum permittivity and the effective permittivity of the surrounding medium. One of the most attractive properties of 2D plasmons is that their frequency can be tuned over a wide range by changing the electron density. This property finds different applications involving the detection and generation of terahertz radiation~\cite{Shur:1993, Knap:2002, Shaner:2005, Aizin:2006, Muravev:2012}.

The geometry of a 2DES significantly influences the physical properties of plasmon modes excited in a sample. Although the properties of 2D plasmons in an unbounded two-dimensional layer have been well-studied, considering confined structures makes the situation exceedingly more complex due to the non-locality of Maxwell’s equations. In this case, determining the plasmonic modes entails a subtle interplay of approximate analytical methods, computer simulations, and symmetry analysis. The two most studied 2DES configurations are the disk~\cite{Allen:1983} and the stripe geometries~\cite{Heitmann:1991, Kukushkin:2005}. For the disk, the fundamental plasmon mode is described by Eq.~(\ref{2D}), with $q \approx 2.4/d$, $\varepsilon(q) = \overline \varepsilon = (\varepsilon + 1)/2$, and $\varepsilon$ denoting the dielectric permittivity of the semiconductor substrate hosting the 2DES. For the stripe shape, the situation is a little more intricate.  For narrow stripes, the plasmon dispersion in Eq.~(\ref{2D}) becomes linear due to strong modification of the Coulomb interaction between the charge fluctuations, which can be accounted for by the effective dielectric function $\varepsilon (q) \approx \overline \varepsilon \, \pi/ Wq$.

The sample with a square geometry can be readily produced by cutting off the wafer. Nonetheless, despite the simplicity of sample fabrication, plasmon modes in a square have been studied the least~\cite{Koppens:2016, Sydoruk:2021}. Therefore, in the present work, we experimentally investigate the fundamental and harmonic plasmon modes in a single 2DES square.  Interestingly, we find the fundamental mode frequency in the square to be practically identical to that in a disk of the same lateral size. However, unlike the disk configuration, magnetoplasmon modes in the square exhibit strong hybridization with each other. In our study, these features were observed experimentally, reproduced in the numerical simulations, and justified by the developed analytic theory.

\section{Samples and experimental method}

The experiments were conducted on a high-quality GaAs/AlGaAs heterostructure, with a $35$~nm wide quantum well at the depth of $160$~nm below the crystal surface. The 2DES had a fixed transport electron mobility $\mu=5\times10^6~\text{cm}^2/\text{V$\cdot$s}$ at $T=4.2$~K, while the electron density $n_s  = (1.0 - 2.2) \times10^{11}~\text{cm}^{-2}$ for different samples. The given heterostructure was used to fabricate several square-shaped 2DESs with the side dimensions of $a=0.5$ and $1.0$~mm. To compare experimental results with a circular geometry, a few disks of equivalent sizes, with diameters $d=0.5$ and $1.0$~mm, were made from the same wafer. The samples were produced either lithographically or by cutting the wafer into square pieces. In order to match the microwave impedance, the sample was placed into a standard $50$ $\Omega$ SMA termination placed at the end of a coaxial cable (inset to Fig.~\ref{fig1}). Such a setup allowed us to measure the 2DES microwave absorption by sweeping the frequency over a wide range of up to $30$~GHz. The microwave absorption was registered employing a noninvasive optical technique based on the high sensitivity of the 2DES luminescence spectrum to its resonant heating~\cite{Kukushkin_2002}. For this purpose, the excitation radiation from a stabilized semiconductor laser of the wavelength $\lambda=780$~nm was delivered to the sample through a $0.4$~mm fiber. The same fiber was used to collect the reflected luminescence signal, which was then fed to the input of the spectrometer with a built-in CCD (charge-coupled device) camera. The samples were placed inside a liquid-helium cryostat at the center of the superconducting coil. All measurements were taken at the temperature $T=4.2$~K, with the magnetic field $B=0 - 2$~T applied perpendicular to the sample surface. 

\section{Experimental results}

\begin{figure}[!t]
\includegraphics[width=\linewidth]{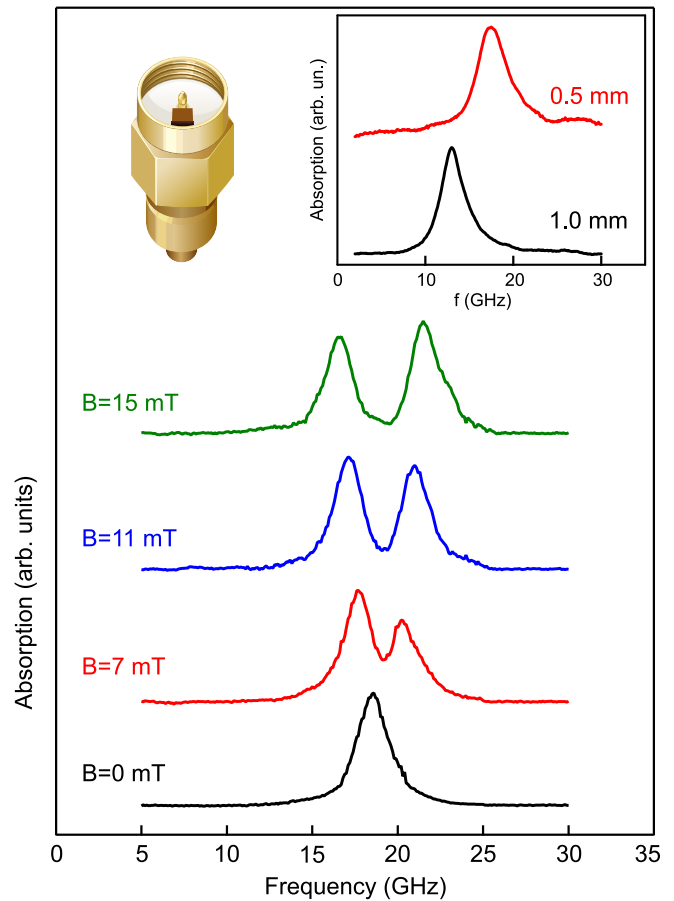} 
\caption{Microwave absorption as a function of frequency, measured at $0$, $7$, $11$ and $15$~mT for the square-shaped 2DES with $a=1$~mm. The 2DES has the density $n_s = 1.8 \times10^{11}~\text{cm}^{-2}$. The left inset shows the schematic drawing of the sample placed in a coaxial SMA termination. The right inset compares the fundamental plasmon resonances measured for two 2DES squares of different sizes with the density $n_s = 1.1 \times10^{11}~\text{cm}^{-2}$.}
\label{fig1}
\end{figure}

\begin{figure}[!t]
\includegraphics[width=\linewidth]{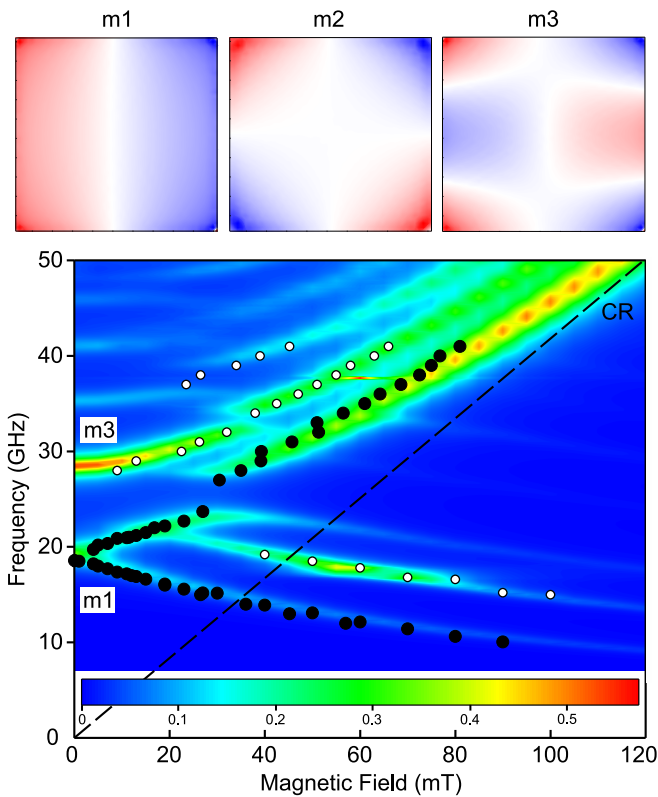} 
\caption{Magnetodispersion of the plasma resonances observed in the square-shaped 2DES, $1 \times 1$~mm$^2$ in size. The black and white dots are the resonance positions determined from the experiment. The dashed line is the CR frequency $\omega_c=e B / m^{\ast}$. The color plots display the numerical simulation results for the absorption cross-section. The upper insets illustrate the amplitude distribution of charge oscillations across the samples, simulated at $B=0$~T.}
\label{fig2}
\end{figure}

Figure~\ref{fig1} displays the dependencies of microwave absorption on the frequency of microwave excitation obtained for a $1 \times 1$~mm$^2$ square-shaped 2DES with $n_s = 1.8 \times10^{11}~\text{cm}^{-2}$. The curves recorded for different magnetic field values are vertically offset for clarity. The absorption spectrum clearly indicates a single resonance at $B=0$~T, attributed to the excitation of the fundamental plasmon mode. The inset to Fig.~\ref{fig1} shows that the frequency of the plasmon resonance scales inversely proportionally to the size of the square, in agreement with the plasmon spectrum (\ref{2D}). In the presence of the magnetic field, the plasmon resonance splits into two modes. To examine the nature of the modes further, we include their magnetodispersion in Fig.~\ref{fig2}, plotted in black dots. In the figure, the low-frequency branch corresponds to the excitation of the edge magnetoplasmon (EMP)~\cite{Allen:1983, Mast:1985, Glattli:1985, Volkov:1988, Fetter:1985}. In a qualitative sense, this mode originates from the skipping-orbit collective motion of electrons along the edge of the 2DES. The high-frequency branch in Fig.~\ref{fig2} refers to the excitation of the cyclotron magnetoplasmon mode arising from the collective motion of electrons in cyclotron orbits throughout the entire area of the 2DES~\cite{Theis:1977}. In the limit of a large magnetic field, the frequency of the cyclotron magnetoplasmon mode tends to that of the cyclotron resonance (CR) $\omega_c=e B / m^{\ast}$, indicated by the dashed line in Fig.~\ref{fig2}. The harmonics of plasma resonances are also observed in the absorption, denoted in the figure by the white dots.

\begin{figure}[!t]
\includegraphics[width=\linewidth]{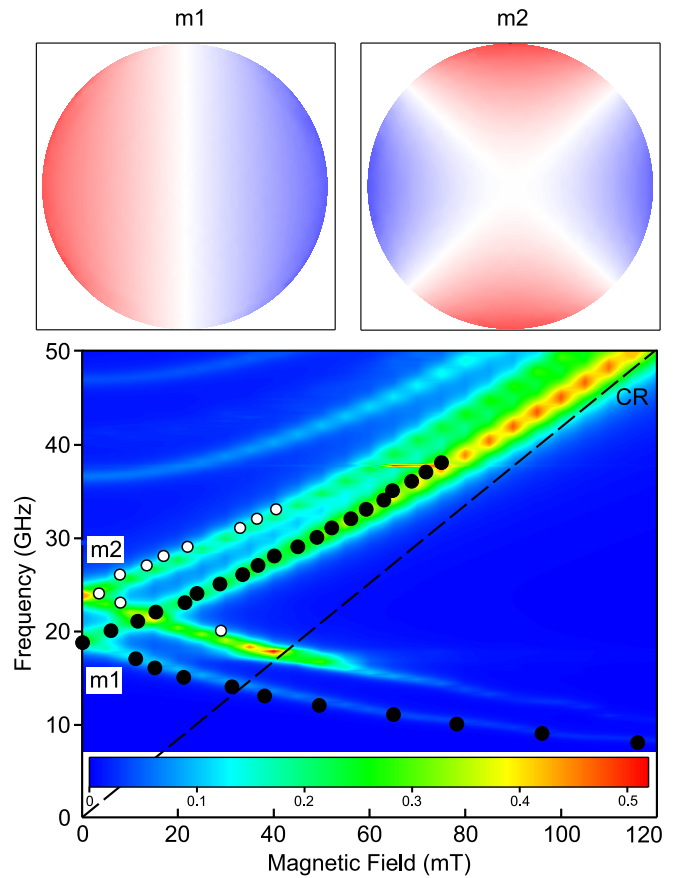} 
\caption{Magnetodispersion of the plasma resonances observed in the disk-shaped 2DES of diameter $d=1$~mm. The black and white dots correspond to the resonances determined from the experiment. The color plots display the numerical simulation results for the absorption cross-section. The upper insets illustrate the amplitude distribution of charge oscillations across the samples, simulated at $B=0$~T.}
\label{fig3}
\end{figure}

It is instructive to compare magnetoplasmon modes excited in the square and disk geometries. The transition from the disk to square geometry causes a significant reduction in symmetry, which is expected to influence the properties of the magnetoplasma excitations. Figure~\ref{fig3} shows the magnetodispersion measured for the disk of diameter $d=1$~mm and electron density $n_s = 1.9 \times10^{11}~\text{cm}^{-2}$. We observe a series of plasmon resonances, each of which splits into the edge and cyclotron magnetoplasmons in the presence of the magnetic field~\cite{Allen:1983}. In a disk, magnetoplasmon modes can be conveniently classified by the azimuthal $m$ and radial $l$ indices, referring to the number of plasma wave nodes along the 2DES disk perimeter and radius. Thus, for the fundamental plasmon mode, $m=1$ and $l=1$, while for the second harmonic, $m=2$ and $l=1$. We note that due to the disk symmetry, magnetoplasmon modes with different indices do not interact with each other. Therefore, there is no anticrossing between the $m=1$ and $m=2$ modes in Fig.~\ref{fig3}. 

To understand the nature of the observed plasmon modes more fully, we performed numerical simulations of the electromagnetic fields using the Microwave Studio package (for details we refer to Supplementary Material I~\cite{Supplementary}). The results are displayed in Fig.~\ref{fig2} and Fig.~\ref{fig3} as color plots, representing the absorption cross-section in arbitrary units. The figures demonstrate a good agreement between the absorption cross-section maxima and the experimental data for the plasmon resonance positions. The upper insets in Fig.~\ref{fig2} and Fig.~\ref{fig3} visualize the charge distributions across the samples calculated for $m=1, 2, 3$ modes in the square and $m=1, 2$ modes in the disk, at $B=0$~T. These distributions help to identify the modes and understand their physical nature. Thus, the simulation provides the key to shedding light on the physics of each plasmon mode.

Comparing the resultant magnetodispersions for the square and the disk leads to several interesting conclusions. First, the resonant frequencies of the fundamental plasmon modes obtained in the square and disk samples appear to be identical $f_p=19.0$~GHz. Second, a strong coupling between the fundamental $m=+1$ and $m=-3$ magnetoplasmon modes is evident in the square sample (Fig.~\ref{fig2}), manifested as anticrossing at $24$~GHz. Simulations also predict similar features for higher frequency harmonics. Notably, for the disk, no coupling is observed between the magnetoplasmon modes (Fig.~\ref{fig2}) due to the high symmetry of the disk geometry, as discussed in the following theoretical section of the paper.

\section{Theoretical considerations}

While purely numerical electromagnetic simulations can readily reproduce the main features of the experiment, they cannot reveal the fundamental reasons for the presence or absence of magnetically-induced mode interaction. The situation calls upon the analogies with quantum mechanics and electron Landau levels in quantum wells of different symmetry. Fortunately, drawing on such an analogy, we can predict the magnetodispersion of plasmons analytically.

To this end, we follow our recent works~\cite{Petrov:2019,Petrov:2020,Petrov:2022} and recast the system of electrostatic equations and electrons' equations of motion (the continuity equation and the Euler equation) in an operator form:
	\begin{equation}
		\label{eq-Operator}
		(\hat{\Omega}+\hat{\Omega}_{c})\mathbf{\Phi} = \Omega\mathbf{\Phi},
	\end{equation}
where $\Omega$ is the plasmon frequency and $\mathbf \Phi = (\delta \rho, \delta j_x, \delta j_y)$ is the plasmon 'state vector' --- a function of the charge density variation $\delta \rho$ and the current variation $\delta {\bf j} = (\delta j_x, \delta j_y)$. The $\hat{\Omega}$-operator describes the motion of electrons {\it unperturbed} by the magnetic field, as follows:
	\begin{equation}
		\label{eq-def-Omega}
		\hat{\Omega} = 
		-i\begin{pmatrix} 
			0 & \partial_x & \partial_y \\
			\frac{e^2n_0}{m}\partial_x\mathcal G[\cdot] & 0 & 0 \\
			\frac{e^2n_0}{m}\partial_y\mathcal G[\cdot] & 0 & 0 \\
		\end{pmatrix}
	\end{equation}
where $\mathcal G[\cdot]$ is the Green's operator defined according to the Poisson's equation:
\begin{equation}
	\delta	\varphi = \mathcal G[\delta\rho] = \int G(x,x',y,y') \delta \rho(x',y')\,dx'dy',
\end{equation}
where $G(x,x',y,y')$ is the Green's function of the electrostatic problem. The action of magnetic field is described by the $\hat{\Omega}_c$-operator:
	\begin{equation}
		\label{eq-def-Omega-c}
		\hat{\Omega}_c = 
		-i\Omega_c\begin{pmatrix} 
			0 & 0 & 0 \\
			0 & 0 & 1 \\
			0 & -1 & 0 \\
		\end{pmatrix}.
	\end{equation}
Physically, the magnetic field acts to rotate the electron trajectory as implied by the form of $\hat{\Omega}_c$, which couples the $x$- and $y$-components of the electric current.

Furthermore, we aim to develop a basis-expansion approach to the problem of plasmon magnetodispersions, akin to the one used to describe the Seeman effect in atoms. Wherefore, applying $\hat{H}$-operator to both sides of Eq.~(\ref{eq-Operator}), we arrive at:
	\begin{equation}
		\label{eq-Operator-Generalized}
		\hat{H}(\hat{\Omega}+\hat{\Omega}_{c})\mathbf{\Phi} = \Omega\hat{H}\mathbf{\Phi},
	\end{equation}
where
\begin{equation}
		\hat{H} = \frac{1}{2}
		\begin{pmatrix}
			\,\mathcal G[\cdot] & 0 & 0 \\
			0 & m/n_0 & 0 \\
			0 & 0 & m/n_0\\
		\end{pmatrix}.
\end{equation}
Physically, $\hat{H}$ can be called 'plasmon Hamiltonian' as its diagonal elements yield the sum of kinetic and potential energy densities in a wave:
\begin{equation}
		 \bra{\mathbf\Phi}\hat{H}\ket{\mathbf \Phi} = \frac{1}{2}\int  d^2 \mathbf r \left[ \frac{ m\mathbf \delta {\bf j}^2}{n_0}  +  \delta \varphi \delta \rho \right].
\end{equation}

The subsequent steps naturally follow the basis-function methods of quantum mechanics. We choose the set of functions, $\mathbf\Phi_{\alpha}$, that solves the spectral problem in the absence of the magnetic field, resulting in the eigenfrequencies $\Omega_\alpha$. In the presence of the magnetic field, the plasmon state vector becomes a linear summation $\mathbf\Phi = \sum_\alpha c_{\alpha}\mathbf\Phi_{\alpha}$. Hence, the expansion coefficients $c_\alpha$ along with the perturbation frequencies $\Omega$ can be found from the linear system of equations:
\begin{equation}
\label{eq-Matrix-Generalized}
   \sum\limits_\beta\left[ (\Omega_\alpha - \Omega)\delta_{\alpha\beta} + \frac{(\hat{H}\hat{\Omega}_c)_{\alpha\beta}}{H_{\alpha\alpha}}\right]c_\beta = 0,
\end{equation}
with the matrix elements $(\hat{H}\hat{\Omega}_c)_{\alpha\beta}$ and $H_{\alpha\alpha}$ given by
\begin{equation}
    \label{eq-Matrix-Elements}
    \begin{split}
       (\hat{H}\hat{\Omega}_c)_{\alpha\beta} & = 
        -\frac{i\Omega_c}{2n_0}\int\limits d^2\mathbf r\left(\delta j_{x,\beta}^*\delta j_{y,\alpha} - \delta j_{y,\beta}^*\delta j_{x,\alpha}\right), \\
       \hat{H}_{\alpha\alpha} & =  \frac{1}{2}\int  d^2 \mathbf r \left[ \frac{ m\mathbf \delta {\bf j}_\alpha^2}{n_0}  +  \delta \varphi_\alpha \delta \rho_\alpha \right].
    \end{split}
\end{equation}

The eigen-frequencies and eigen-modes in magnetized 2DES are now readily found from Eq.~(\ref{eq-Matrix-Generalized}), as soon as basis mode profiles at $B=0$ are known from simulations. An even more transparent estimate can be obtained if we approximate the basis functions with harmonic functions with properly chosen wave vector and phase (for their explicit form please refer to Supplementary Material~\cite{Supplementary}).

Further, we restrict our basis to four lowest modes which we label as $\ket{X_1 Y_0}$, $\ket{X_0 Y_1}$, $\ket{X_2 Y_1}$ and $\ket{X_1 Y_2}$, where $X_\alpha$ ($Y_\alpha$) stands for a mode having $\alpha$ zeros along the $x$ ($y$) axis. As a result, we arrive at the following system
\begin{equation}
\label{eq-matrix-dispersion}
\left(
\begin{array}{c|c}
M_{11} & M_{12} \\ \hline
M_{21} & M_{22} \\
\end{array}
\right)
\mathbf{c} = \mathbf{0},
\end{equation}
where
\begin{equation}
\begin{split}
        M_{11} & = \begin{pmatrix} 
			\Omega_{10}-\Omega  & -i\Omega_c a_1^2 \\
			i\Omega_c a_1^2 & \Omega_{01} - \Omega  \\ 
    \end{pmatrix}; \\
     M_{12}& = \begin{pmatrix} 
			 -i\Omega_c a_1 a_3 & 0 \\
			 0 & i\Omega_c a_1 a_3 
    \end{pmatrix}; \\
     M_{21}& = \begin{pmatrix} 
			 i\frac25 \Omega_c a_1 a_3 & 0 \\
			 0 & -i\frac25\Omega_c a_1 a_3 
    \end{pmatrix}; \\
    M_{22} & = \begin{pmatrix} 
			 \Omega_{21}-\Omega & -6 i\Omega_c a_3^2 \\
			 6 i\Omega_c a_3^2 & \Omega_{12}-\Omega 
    \end{pmatrix}; \\
    \mathbf{c} & = (c_{10},\;c_{01}\;,c_{21}\;, c_{12})^T,
\end{split}
\end{equation}
and $a_k = 2/(\pi k)$. 

The elements $M_{11}$ and $M_{22}$ describe the magnetic-field-induced splitting of the $\ket{X_0Y_1}$, $\ket{X_1Y_0}$, and $\ket{X_2Y_1}$, $\ket{X_1Y_2}$ modes, respectively. In particular, setting the determinant of $M_{11}$ to $0$ allows us to reproduce the experimentally observed lower mode slope $d\omega/d\omega_c = \pm 4/\pi^2$ in the weak magnetic fields. At the same time, the upper mode slope of $\pm 8/3\pi^2$ is calculated from $det\,M_{22}=0$ to be significantly less, which is also consistent with the experiment. 

The off-diagonal elements $M_{12}$ and $M_{21}$ describe the hybridization between the lower and upper pairs of modes. Crucially, in the disk geometry, $M_{12} = M_{21} \equiv 0$, and the hybridization is prohibited. The reason for such behavior is due to the fundamental difference between the disk and square symmetries. Indeed, the square geometry simultaneously hosts both even and odd modes, with an integer or half-integer number of periods on the square side, accordingly. However, the disk geometry supports only even modes with an integer number of periods along the circumference. That is why the plasmon modes in a disk do not hybridize --- any integral of an even function taken over its period vanishes. By contrast, in the square geometry, even and odd modes interact with each other, and the resulting integrals of odd functions do not vanish.

\begin{figure}[!t]
\includegraphics[width=\linewidth]{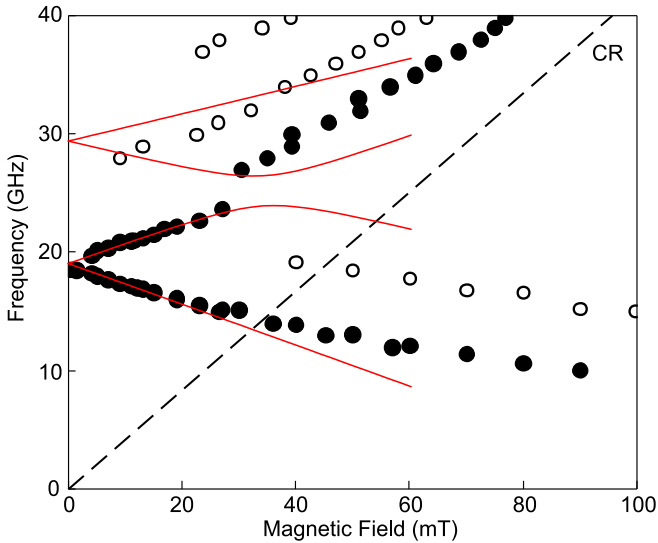} 
\caption{Analytical dispersion (red curves) compared to experimental data (black dots and emplty circles) obtained for the plasmons in a $1 \times 1$~mm$^2$ 2DES, at $n_s = 1.8\cdot 10^{11}\,\text{cm}^{-2}$.}
\label{fig4}
\end{figure}

To obtain the dispersion dependence $\Omega(\Omega_c)$, we set the determinant of the matrix in (\ref{eq-matrix-dispersion}) to zero. In Fig.\ref{fig4}, we compare the analytic solutions (solid red curves) with the experimental results (black dots and empty circles). The figure demonstrates a close agreement between theory and experiment up to the magnetic field level of $50$~mT, where our analytical model accurately reproduces the key dispersion features, such as the mode anticrossing and the correct dispersion slope at low magnetic fields.

In conclusion, we experimentally investigate the microwave absorption spectra of a square of two-dimensional electrons. We identify the resonant plasmon modes. A comparison between the plasmon spectra in the square and disk geometries reveals several fundamental differences arising from the symmetry particularities of the electromagnetic fields. We find the frequencies of the fundamental plasmon modes in square and disk-shaped samples to be practically identical. However, unlike the disk configuration, the square geometry leads to a strong interaction between the fundamental and harmonic plasmon modes in the presence of the magnetic field. We develop the theory based on the plasmonic perturbation approach that accurately describes our experimental findings. As the square 2DES geometry is widespread in modern electronics, the results reported in this paper can be applied in construction of 2D plasmonic devices such as terahertz detectors, mixers, emitters, and ﬁeld enhancement structures.

The authors gratefully acknowledge the financial support from the Russian Science Foundation (Grant No.~18-72-10072 for experiment. Theoretical work of AP, DM and DS was supported by grant No.~MK-1035.2021.4 of the President of Russian Federation).

\end{document}